\newcommand{\slh}{\!\!\!\slash}

\documentclass[twocolumn,showpacs,preprintnumbers,amsmath,amssymb,letterpaper]{revtex4}

\begin{document}

\title{
Detecting topological orders through continuous quantum phase transition
}
\author{Ying Ran}
\author{Xiao-Gang Wen}
\homepage{http://dao.mit.edu/~wen}
\affiliation{Department of Physics, Massachusetts Institute of Technology,
Cambridge, Massachusetts 02139}
\date{June 2005}

\begin{abstract}
We study a continuous quantum phase transition that breaks a $Z_2$ symmetry.
We show that the transition is described by a new critical point which does not
belong to the Ising universality class, despite the presence of well defined
symmetry breaking order parameter.  The new critical point arises since the
transition not only break the $Z_2$ symmetry, it also changes the
topological/quantum order in the two phases across the transition.  We show
that the new critical point can be identified in experiments by measuring
critical exponents.  So  measuring critical exponents and identifying new
critical points is a way to detect new topological phases and a way to measure
topological/quantum orders in those phases.
\end{abstract}

\maketitle

For a long time, Landau symmetry-breaking theory \cite{L3726,LanL58} is
believed to be the theory that describes all possible phases and phase
transitions.  The Ginzburg-Landau theory \cite{GL5064} based on order
parameters and long rang order became the standard theory for all
kinds of continuous phase transitions.

However, after the discovery of fractional quantum Hall (FQH)
effect,\cite{TSG8259} people realized that different FQH states all have the
same symmetry. So the order in FQH states cannot be described by the Landau's
symmetry breaking theory.  The new order is called topological order.
\cite{Wrig,Wtoprev} Topological order is new since it has nothing to do with
symmetry breaking, long range correlation, or local order parameters.  None of
the usual tools that we used to describe a symmetry breaking phase applies to
topological order.  Despite this, topological order is not an empty concept
since it can be described by a new set of tools, such as the number of
degenerate ground states \cite{HR8529,WNtop}, quasiparticle statistics
\cite{ASW8422}, and edge states \cite{H8285,Wedgerev,Wtoprev}.

The existence of topological orders has consequences on our understanding of
continuous phase transitions.  If there exists phases that are not described by
symmetry breaking, then it is reasonable to guess that there exist continuous
phase transitions that are not described by changes of symmetries and the
associated order parameters.  Indeed continuous phase transitions exist between
two phases with the \emph{same} symmetry
\cite{WWtran,CFW9349,SMF9945,RG0067,Wctpt} and between two phases with the
\emph{incompatible}\footnote{If the symmetry group of one phase is not a
subgroup of the other phase and vice versa, then the two phases are said to have
incompatible symmetries.  According to Landau's symmetry breaking theory, two
phases with incompatible symmetries cannot have continuous phase transition
between them.} symmetries.\cite{SVB0490}.  In this paper, we will show that
even some symmetry breaking continuous phase transitions are beyond Landau's
symmetry breaking paradigm in the sense that critical properties of the
transition are not described by fluctuating symmetry breaking order parameters
and not described by Ginzburg-Landau effective theories.  As a results, the
critical exponents of those symmetry breaking transitions are different for
those obtained from Ginzburg-Landau theory.

Why some symmetry breaking transitions give rise to new class of critical
points?  One reason is  that those transitions not only change the symmetry of
the states, they also changes the topological/quantum order in the states. So
the appearance of the new critical points, in many cases, imply the appearance
of new state of matter with non-trivial topological/quantum orders.  It is known
that frustrated spin systems on Kagome or pyrochlore lattices contain many
different quantum phases.  Those different quantum phases in general contain
different spin orders as shown by magnetic susceptibility measurements.  So one
naturally assume those spin ordered phases are described by symmetry breaking,
and the continuous transition between those phases are symmetry breaking
transitions described by Ginzburg-Landau theory.  The main message of this
paper is that those $T=0$ spin ordered phases may contain additional
topological orders and represent new states of matter.  The additional
topological orders can be detected by measuring critical exponents at
continuous quantum transition points between those $T=0$ quantum phases (even
when the continuous transitions are symmetry breaking transitions).  If the
measured critical exponents do not agree with the those values obtained from
Ginzburg-Landau theory, then the transition point will be a new quantum
critical point and the two phases separated by the phase transition will
contain non-trivial topological orders.


But why zero temperature is important?
Can we find new states of matter and new continuous phase transitions at finite
temperatures? The answer is yes, but it is more difficult to find new states of
matter and new continuous transitions at finite temperatures. This is because
most of the known new states of matter are due to string-net condensations.
\cite{LWstrnet,Wen04} String-net condensations and the continuous phase
transition between different string-net condensed states only exist at zero
temperature. So it is much easier to find new states of matter and new
continuous transitions at zero temperature.


To illustrate our point that a continuous $T=0$
symmetry breaking phase transition
can have a new set of critical exponents beyond Ginzburg-Landau symmetry
breaking theory, we consider a frustrated spin-1/2 model on square lattice:
\begin{align}
\label{heisenberg}
H&=\sum_{\<\v i\v j\>}J_{\v i\v j}\,\mathbf{S}_{\v i} \cdot \mathbf{S}_{\v j}
\end{align}
where the coupling between the nearest neighbors is $J_1$ and between the
second nearest neighbors is $J_2$.  We have used $SU(2)$ slave-boson theory
\cite{AZH8845,DFM8826} to study the possible spin liquid phases of the above
frustrated spin model.  Under mean-field approximation, we find a continuous
phase transition between the two spin liquids at $J_2/J_1\approx 0.52$.
\cite{WWZcsp,RWtoap}

The spin liquid phase for $J_2< 0.52 J_1$ is called $SU(2)$-linear spin liquid
(or $\pi$-flux phase) \cite{AZH8845,DFM8826,Wqoslpub} whose low energy
effective field theory is a $SU(2)$ gauge theory couple to massless Dirac
fermions \cite{RWtoap}
\begin{align}
\label{su2lin}
L&=\sum_{i=1}^{N}
\bar{\psi}_{ai}\left( \partial_{\mu}-ia_\mu^l \tau^l_{ab} \right)\gamma_{\mu}\psi_{bi}
+\frac{1}{4g^2} f^l_{\mu\nu}f^l_{\mu\nu}
\end{align}
where $\psi_{ai}$ is four-component Dirac fermion field, $a=1,2$,
$i=1,2,...,N$, and $N=1$.  $\ga^\mu$, $\mu=0,1,2,3,5$, are Dirac matrices and
the summation of $\mu$ run through $0,1,2$.  $\tau^l$, $l=1,2,3$, are Pauli
matrices and $a^l_{\mu}$ is $SU(2)$ gauge field.

The spin liquid phase for $J_2> 0.52 J_1$ is called chiral spin liquid
\cite{WWZcsp} whose low energy effective field theory is a $SU(2)$ gauge theory
couple to fermions with a chiral mass \cite{WWZcsp,RWtoap}
\begin{align}
\label{csp}
L&=\sum_{i=1}^{N}
\bar{\psi}_{ai}\left( \partial_{\mu}-ia_\mu^l \tau^l_{ab} \right)\gamma_{\mu}\psi_{bi}
+m \bar{\psi}_{ai}(i\ga^3 \ga^5){\psi}_{ai}
\nonumber\\
&\ \ +\frac{ 1}{4g^2} f^l_{\mu\nu}f^l_{\mu\nu}
\end{align}
The effective field theory that describes the transition
connects \eq{su2lin} and \eq{csp} and
is given by \cite{RWtoap}
\begin{align}
\label{su2csptran}
L=&\sum_{i=1}^{N}\bar{\psi}_i\left( \partial_{\mu}-i a_\mu^l \tau^l \right)\gamma_{\mu}\psi_i+\frac{1}{4g^2} 
f^l_{\mu\nu}f^l_{\mu\nu}\notag\\
&+\sigma \bar{\psi}[i\gamma_3\gamma_5]\psi+\frac{1}{2\rho^2}(\partial_{\mu}\sigma)^2+V(\sigma)
\end{align}
where $V(\si)=V(-\si)$.

\begin{figure}
\hspace{0.05\textwidth}
\centerline{
\includegraphics[scale=0.35]{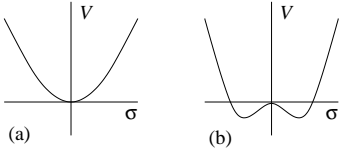}
}
\caption{The behavior of potential $V(\sigma)$ (a) before  and (b)
after the phase transition from the $SU(2)$-linear phase to 
the chiral spin phase.}
\label{Z2tran}
\end{figure}

We note that the $SU(2)$-linear state does not break any symmetry and the
chiral spin state breaks the time reversal and parity symmetry.
The real $\si$ field is the order parameter of the symmetry breaking.
The potential $V(\si)$ controls the phase transition (see Fig. \ref{Z2tran}).
$\si=0$ gives rise to the $SU(2)$-linear state and $\si\neq 0$ gives rise
to the symmetry breaking chiral spin state.
The order parameter $\si$ is related to the following combination of 
physical spin operators: 
$\si\propto \v S_{\v i}\cdot(\v S_{\v i+\v x}\times \v S_{\v i+\v y})$
\cite{WWZcsp}


The transition between the $SU(2)$-linear and the chiral spin state 
is a $Z_2$
symmetry breaking transition. So we may expect the critical
point to belong to the universality class of 3D Ising model.  For 3D Ising
model, the order parameter has scaling dimension $[\si]^\text{Ising}\approx0.51$ (or a
correlation $\<\si(x)\si(0)\>=x^{-2[\si]^\text{Ising}}$) at the critical point.
One may conclude that, at the transition point between the $SU(2)$-linear and
the chiral spin state, the order parameter $\v S_{\v i}\cdot(\v S_{\v i+\v
x}\times \v S_{\v i+\v y})$ also has the correlation
\begin{equation*} 
\<\v S_{\v i}\cdot(\v S_{\v i+\v x}\times \v S_{\v i+\v y})
\v S_{\v j}\cdot(\v S_{\v j+\v x}\times \v S_{\v j+\v y})\> \propto 
|\v i-\v j|^{-2[\si]^\text{Ising}} 
\end{equation*} 
In fact the above guess is incorrect.  For our case, even though the transition
breaks a $Z_2$ symmetry and has a well defined $Z_2$ order parameter, the
critical point does not belong to the 3D Ising class.

Why the $SU(2)$-linear state to the chiral spin state transition belongs to a
new universality class? The reason is that, at the critical point, not only the
fluctuations of the order parameter $\si$ give rise to gapless excitations, the
fermion field $\psi$ and the $SU(2)$ gauge field $a^l_\mu$ also give rise to
gapless excitations.  Had all gapless excitations come from the order parameter
$\si$, then the transition would be belong to the 3D Ising class.  So the key
to understand the existence of the new critical point is to understand why
$\psi$ and $a^l_\mu$ can give rise to gapless excitations.

\begin{figure}
\hspace{0.05\textwidth}
\centerline{
\includegraphics[scale=0.30]{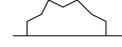}
}
\caption{
The fermion self energy due to the gauge interaction. The solid lines
represent the fermion propagator and wiggled line gauge propagator.
}
\label{selfE}
\end{figure}

At first sight, one may expect both $\psi$ and $a^l_\mu$ are gapped due to
their interaction. In fact, the self energy term (see Fig. \ref{selfE})  from
the $SU(2)$ gauge interaction, in general, can generate a fermion mass term
$\del m\bar\psi_i\psi_i$. Once the fermions are gapped, the $SU(2)$ gauge field
is always in confined phase in 1+2 dimensions and the gauge bosons are also
gapped. So in order for the new critical point described by the gapless $\si$,
$\psi$ and $a^l_\mu$ field to exist, we must find the reason that protect the
gaplessness of $\psi$ and $a^l_\mu$.

\begin{figure}[t]
\includegraphics[width=0.4\textwidth]{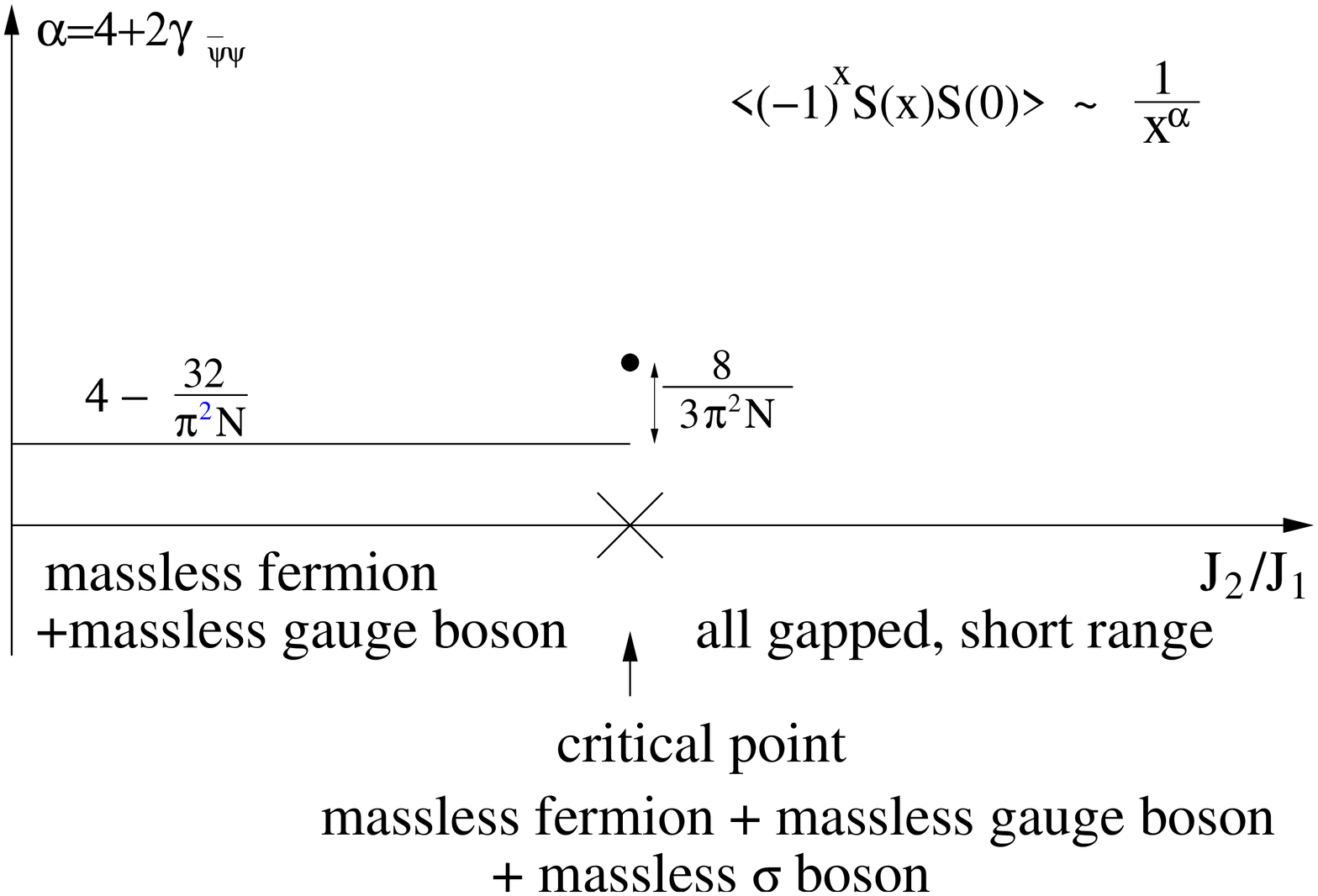}
\caption{Change of scaling dimension of staggered spin-spin correlation function during phase transition.}\label{su2chiraltran}
\end{figure}

To find such a reason, we like to point out that both the $SU(2)$-linear state
and the chiral spin state contain non-trivial quantum or topological orders.
Such quantum/topological orders are characterized by projective symmetry groups
(PSG) which describe symmetry of the effective theory. \cite{Wqoslpub,Wen04}
The PSGs of the $SU(2)$-linear state and the chiral spin state are studied in
detail and will be described in a forthcoming paper. \cite{RWtoap} We find that
it is PSG that protects the gaplessness of $\psi$ and $a^l_\mu$.
\cite{Wqoslpub,WZqoind,RWtoap} In other words, if we regulate the effective
field theory (\ref{su2csptran}) in a way that does not break the symmetry
described by the PSG, then the self energy term in Fig. \ref{selfE} cannot
generate fermion mass. \cite{RWtoap} With the gapless excitations from $\si$,
$\psi$ and $a^l_\mu$, the effective field theory (\ref{su2csptran}) describes a
new critical point at the transition between the $SU(2)$-linear state and the
chiral spin state.

The scaling dimensions of operators are in general difficult to calculate.
However, if we assume the number of the fermion fields $N$ to be large (instead
of $N=1$), then they can be calculated systematically in $1/N$ expansion.
\cite{RWtoap} We find that the order parameter $\si$ has a scaling dimension
$[\si]=1+O(\frac{1}{N})$ at the new critical point, which is 
different from the scaling dimension for the 3D Ising universality class given
by $[\si]^\text{Ising}=0.51$.  Near the transition, there is a diverging
length scale $\xi \propto |t-t_c|^{-\nu}$ where $t$ is a parameter (such as
$J_2/J_1$) that controls the transition. For the new critical point, the
coherent length exponent $\nu$ is found to be $\nu=1+O(\frac{1}{N})$, while for
the 3D Ising universality class $\nu^\text{Ising}\approx 0.63$.

We can also calculate the staggered spin-spin correlations which is easier to
measure.  In  $SU(2)$-linear phase and at the critical point, spins have
algebraic correlations with different exponents. In the chiral spin phase, the
spins have short ranged correlation (see Fig. \ref{su2chiraltran}).

The scaling dimension of staggered spin-spin correlation function $\<(-)^{\v
x}\mathbf{S}(\v x)\mathbf{S}(\v 0)\>$ is calculated by the large-$N$ expansion
of quantum field theory. In our formalism, one can show that the staggered
spin-spin correlation function is just the correlation function of the fermion
mass operator $\<\bar{\psi}\psi(\v x)\bar{\psi}\psi(\v 0)\>$ in the effective
theory Eq.(\ref{su2csptran}). By power counting, the scaling behavior should be
$\<\bar{\psi}\psi(\v x)\bar{\psi}\psi(\v 0)\>=x^{-4}$, but quantum fluctuation
change the it into $\<\bar{\psi}\psi(\v x)\bar{\psi}\psi(\v
0)\>=x^{-4-2\gamma_{\bar{\psi}\psi}}$, where $\gamma_{\bar{\psi}\psi}$ is
called the anomalous dimension of fermion mass operator. 

In the following, we will use the spin correlation as an example to
demonstrate how various correlations are calculated in the large $N$ limit.  It
turned out that the easiest way of calculating $\gamma_{\bar{\psi}\psi}$ is not
to calculate $\<\bar{\psi}\psi(\v x)\bar{\psi}\psi(\v 0)\>$ directly, but to
calculate the correlation function of fermion field $\psi$: $\<\psi(\v
x)\bar{\psi}(\v 0)\>$, and the three-point correlation function
$\<\bar{\psi}\psi(\v x)\bar{\psi}(\v y)\psi(\v 0)\>$. Let us firstly calculate
the staggered spin-spin correlation function in $SU(2)$-linear phase, where the
low energy effective theory is Eq.(\ref{su2lin}). We will do our calculations in Landau gauge. 

\begin{figure}
\hspace{\stretch{1}}
\includegraphics[width=0.13\textwidth]{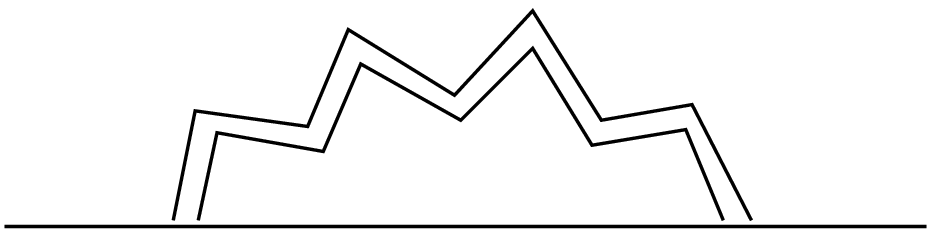}
\hspace{\stretch{3}}
\includegraphics[width=0.22\textwidth]{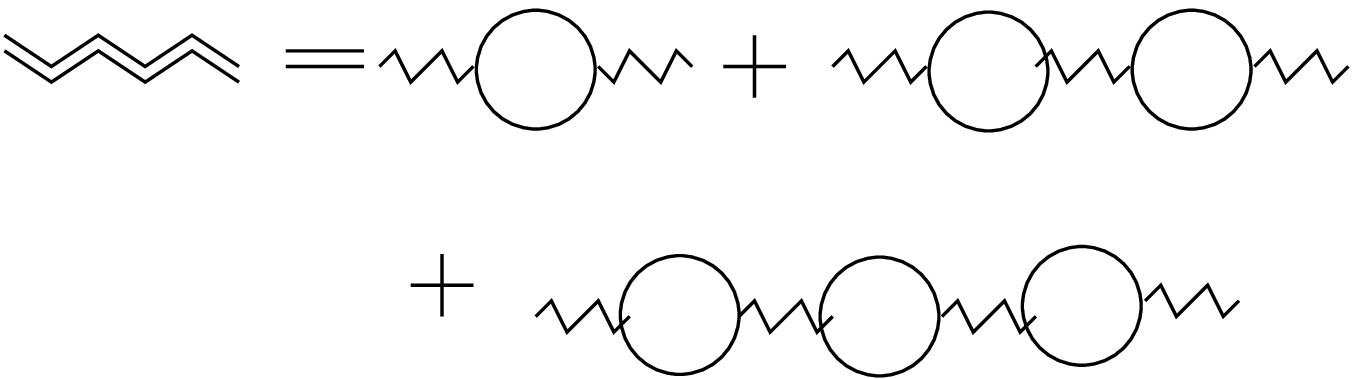}
\hspace{\stretch{1}}
\caption{Gauge dressed fermion propagator at first order of 
$\frac{1}{N}$, where the double wiggled line is the dressed gauge propagator
in the leading order of large $N$ limit.}
\label{gaugedoubleline} 
\end{figure}

In the large-$N$ limit, the gauge field is strongly screened by fermions. To
the leading order of $\frac{1}{N}$, the dressed gauge propagator is shown in
Fig.\ref{gaugedoubleline}. 
The dressed gauge propagator in Landau gauge is found to be:
\begin{align}
&G^{ab}_{\mu\nu,dressed}(k)=\frac{N\delta^{ab}}{16k}\left(\delta_{\mu\nu}-\frac{k_{\mu}k_{\nu}}{k^2}\right)
\end{align}

The fermion correlation to the first order 
in $\frac{1}{N}$, is given by (see Fig.\ref{gaugedoubleline}):
\begin{align}
&S_{dressed}(k)=\frac{-ik\slh}{k^2}(1+\Sigma)\\
p\slh\Sigma&=i\int\frac{dq^3}{(2\pi)^3}\frac{\gamma_{\mu}(-i)(k\slh+q\slh)\gamma_{\nu}}{(k+q)^2}\frac{\frac{3}{4}\cdot16}{Nq}\left(\delta_{\mu\nu}-\frac{q_{\mu}q_{\nu}}{q^2}\right)\notag\\
&=-p\slh\frac{4}{3\pi^2N}\log(\frac{k}{\Lambda})
\end{align}
Therefore the anomalous dimension of $\psi$ is $\gamma_{\psi}=-\frac{1}{2}\frac{4}{\pi^2N}$.

\begin{figure}
\hspace{\stretch{0.5}}
\includegraphics[width=0.08\textwidth]{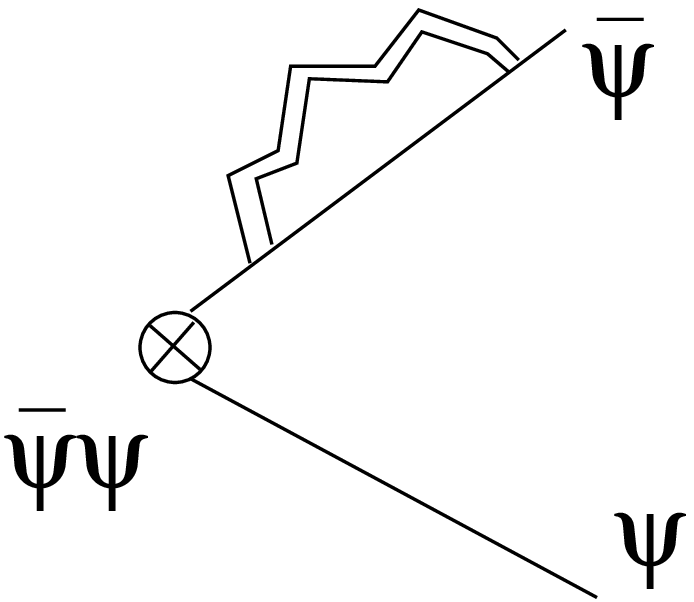}
\hspace{\stretch{1}}
\includegraphics[width=0.08\textwidth]{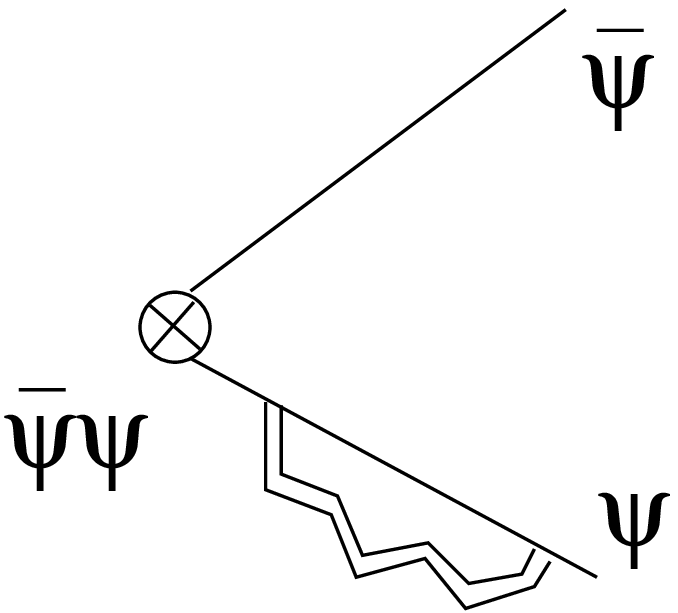}
\hspace{\stretch{1}}
\includegraphics[width=0.08\textwidth]{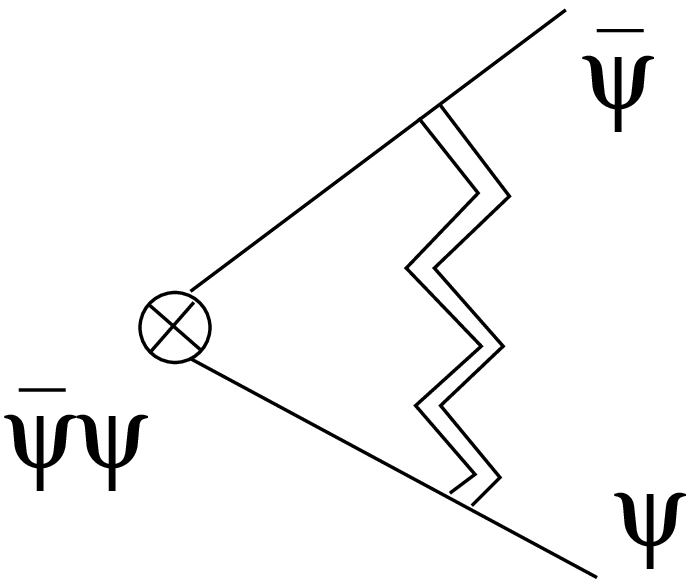}
\hspace{\stretch{0.5}}\\
\hspace{\stretch{0.5}}
A\hspace{\stretch{1}}B\hspace{\stretch{1}}C\hspace{\stretch{0.5}}
\caption{
Gauge dressed three point correlation function at order of 
$\frac{1}{N}$.}\label{gauge_dressed_three_point}
\end{figure}

Then we look at the dressed three-point correlation function
$\<\bar{\psi}\psi(\v x)\bar{\psi}(\v y)\psi(\v 0)\>$ at order of
$\frac{1}{N}$, as shown in Fig.\ref{gauge_dressed_three_point}. Suppose we
fix the momentum of $\bar{\psi}\psi$ to be $2k$, while $\bar{\psi}$ and $\psi$
each carry momentum $k$, then the tree level three point correlation function
will be $G_3(2k,k,k)=\frac{-ik\slh}{k^2}\frac{-i(-k\slh)}{k^2}=\frac{1}{k^2}$.
From the contributions of diagrams in Fig.\ref{gauge_dressed_three_point}, the
dressed three point correlation function is:
\begin{align}
G_{3,dressed}(2k,k,k)=\frac{1}{k^2}\left(1+(A+B+C)\log(\frac{k}{\Lambda})\right)\notag
\end{align}
where $A,B,C$ are the contribution from each corresponding diagram. Actually we
know that $A+B+C=\gamma_{\bar{\psi}\psi}+2\gamma_{\psi}$.

It is easy to see that $A,B$ come from the dressed fermion propagator:
$A=B=2\gamma_{\psi}$. New calculation need to be done for vertex correction in
C. 
\begin{align}
&C\log(\frac{k}{\Lambda})\notag\\
=&\frac{\frac{3}{4}\cdot16}{N}\int\frac{dq^3}{(2\pi)^3}\frac{\gamma_{\mu}(q\slh+k\slh)(q\slh-k\slh)\gamma_{\nu}}{(q+k)^2(q-k)^2q}\left(\delta_{\mu\nu}-\frac{k_{\mu}k_{\nu}}{k^2}\right)\notag\\
=&-\frac{12}{\pi^2N}\log(\frac{k}{\Lambda}).
\end{align}
Thus $\gamma_{\bar{\psi}\psi}=A+B+C-2\gamma_{\psi}=-\frac{16}{\pi^2N}$.

\begin{figure}
\hspace{
\stretch{1}}
\includegraphics[width=0.13\textwidth]{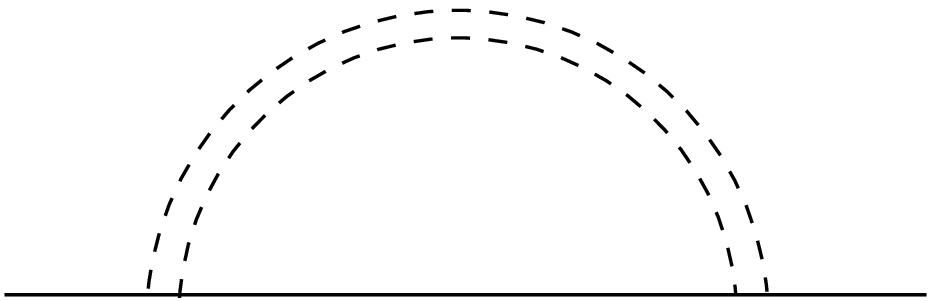}
\hspace{\stretch{3}}
\includegraphics[width=0.22\textwidth]{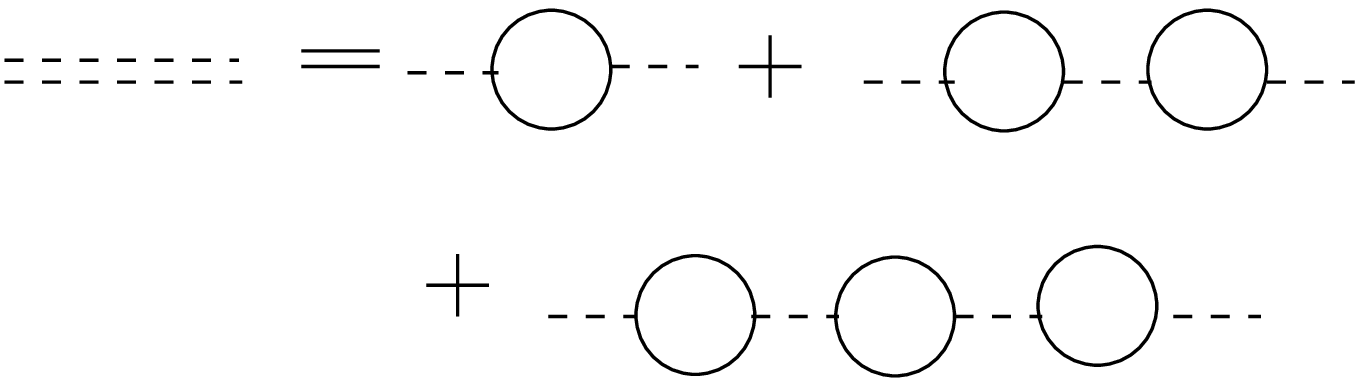}
\caption{The contribution of $\sigma$-boson to fermion propagator at order of $\frac{1}{N}$, where the double dashed line is the dressed $\sigma$-boson propagator at leading order.}\label{boson_dressed_fermion}
\end{figure}

\begin{figure}
\hspace{
\stretch{0.5}}
\includegraphics[width=0.08\textwidth]{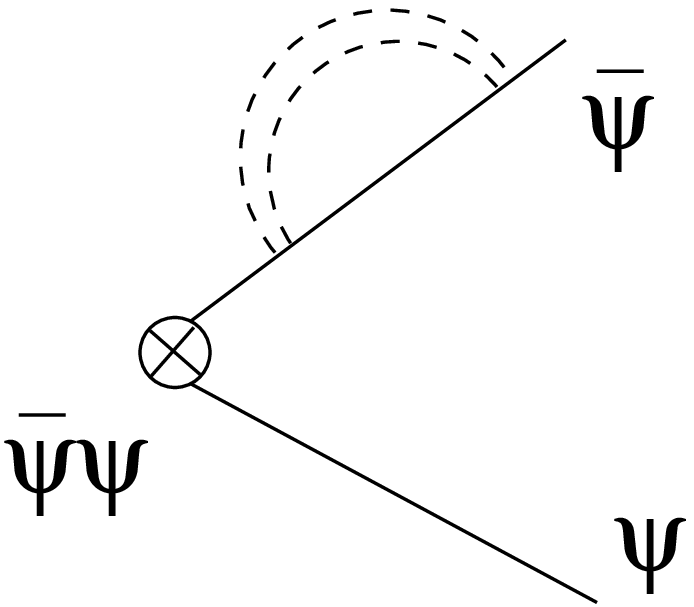}
\hspace{\stretch{1}}
\includegraphics[width=0.08\textwidth]{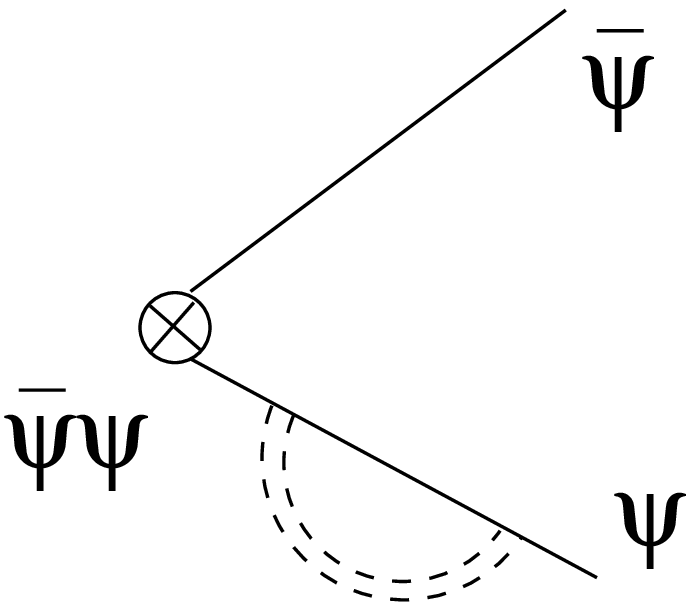}
\hspace{\stretch{1}}
\includegraphics[width=0.08\textwidth]{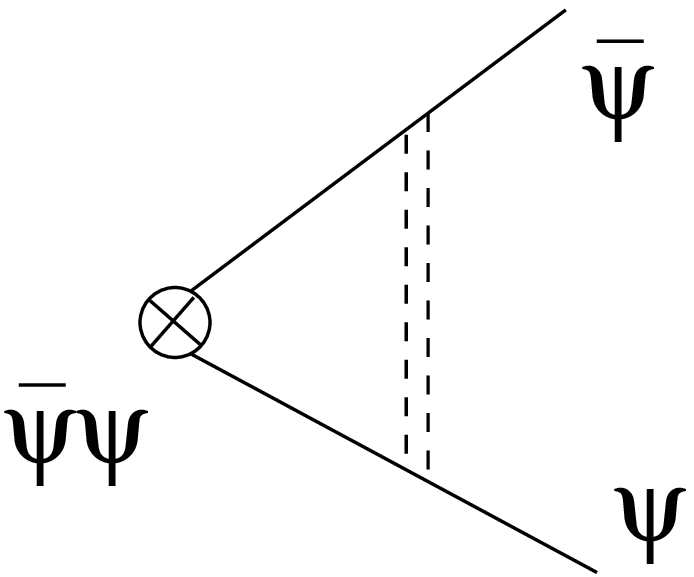}
\hspace{\stretch{0.5}}
\caption{Contributions of $\sigma$-boson to three point correlation function at
order of $\frac{1}{N}$.}
\label{boson_dressed_three_point} 
\end{figure}

We can also calculate the spin-spin correlation function at the critical point
in a similar fashion. The only difference is that the $\sigma$ boson becomes
massless at critical point and contributes to the anomalous dimension of
correlation functions. As shown in Fig.\ref{boson_dressed_fermion} and
Fig.\ref{boson_dressed_three_point}, after similar calculations, we found that
at the critical point,
$\gamma_{\bar{\psi}\psi}=-\frac{16}{\pi^2N}+\frac{4}{3\pi^2N}$, where the
second term comes from contribution of massless $\sigma$-boson.

In this paper, we study a continuous quantum phase transition that breaks a
$Z_2$ symmetry. Despite being a symmetry breaking transition with well defined
order parameter, the transition is described by a new critical point which does
not belong to the Ising universality class.  The new critical point is due to
the fact that the transition not only break the $Z_2$ symmetry, it also changes
the topological/quantum order in the two phases across the transition.  The
additional gapless excitations protected by the PSG change the scaling behavior
and the universality class of the critical point.  
So even a symmetry breaking continuous transition may be described by a new
critical point.  Thus it is very important to measure critical exponents even
for seemingly ordinary symmetry breaking transition.  A new critical point can
be identified by measuring critical exponents and confirming the critical
exponents to be different from those predicted by Ginzburg-Landau theory.
Since a new critical point is usually due to a change in topological/quantum
order at the transition, a discovery of new critical point usually implies a
discovery of new states of matter with non-trivial topological/quantum orders
on the two sides of the transition.

This research was supported by NSF grant No.  DMR-0433632.

\bibliographystyle{apsrev}
\bibliography{/home/wen/bib/wencross,/home/wen/bib/all,/home/wen/bib/misc,/home/wen/bib/publst}

\begin{thebibliography}{25}
\expandafter\ifx\csname natexlab\endcsname\relax\def\natexlab#1{#1}\fi
\expandafter\ifx\csname bibnamefont\endcsname\relax
  \def\bibnamefont#1{#1}\fi
\expandafter\ifx\csname bibfnamefont\endcsname\relax
  \def\bibfnamefont#1{#1}\fi
\expandafter\ifx\csname citenamefont\endcsname\relax
  \def\citenamefont#1{#1}\fi
\expandafter\ifx\csname url\endcsname\relax
  \def\url#1{\texttt{#1}}\fi
\expandafter\ifx\csname urlprefix\endcsname\relax\def\urlprefix{URL }\fi
\providecommand{\bibinfo}[2]{#2}
\providecommand{\eprint}[2][]{\url{#2}}

\bibitem[{\citenamefont{Landau}(1937)}]{L3726}
\bibinfo{author}{\bibfnamefont{L.~D.} \bibnamefont{Landau}},
  \bibinfo{journal}{Phys. Z. Sowjetunion} \textbf{\bibinfo{volume}{11}},
  \bibinfo{pages}{26} (\bibinfo{year}{1937}).

\bibitem[{\citenamefont{Landau and Lifschitz}(1958)}]{LanL58}
\bibinfo{author}{\bibfnamefont{L.~D.} \bibnamefont{Landau}} \bibnamefont{and}
  \bibinfo{author}{\bibfnamefont{E.~M.} \bibnamefont{Lifschitz}},
  \emph{\bibinfo{title}{Statistical Physics - Course of Theoretical Physics Vol
  5}} (\bibinfo{publisher}{Pergamon}, \bibinfo{address}{London},
  \bibinfo{year}{1958}).

\bibitem[{\citenamefont{Ginzburg and Landau}(1950)}]{GL5064}
\bibinfo{author}{\bibfnamefont{V.~L.} \bibnamefont{Ginzburg}} \bibnamefont{and}
  \bibinfo{author}{\bibfnamefont{L.~D.} \bibnamefont{Landau}},
  \bibinfo{journal}{Zh. Ekaper. Teoret. Fiz.} \textbf{\bibinfo{volume}{20}},
  \bibinfo{pages}{1064} (\bibinfo{year}{1950}).

\bibitem[{\citenamefont{Tsui et~al.}(1982)\citenamefont{Tsui, Stormer, and
  Gossard}}]{TSG8259}
\bibinfo{author}{\bibfnamefont{D.~C.} \bibnamefont{Tsui}},
  \bibinfo{author}{\bibfnamefont{H.~L.} \bibnamefont{Stormer}},
  \bibnamefont{and} \bibinfo{author}{\bibfnamefont{A.~C.}
  \bibnamefont{Gossard}}, \bibinfo{journal}{Phys. Rev. Lett.}
  \textbf{\bibinfo{volume}{48}}, \bibinfo{pages}{1559} (\bibinfo{year}{1982}).

\bibitem[{\citenamefont{Wen}(1990)}]{Wrig}
\bibinfo{author}{\bibfnamefont{X.-G.} \bibnamefont{Wen}},
  \bibinfo{journal}{Int. J. Mod. Phys. B} \textbf{\bibinfo{volume}{4}},
  \bibinfo{pages}{239} (\bibinfo{year}{1990}).

\bibitem[{\citenamefont{Wen}(1995)}]{Wtoprev}
\bibinfo{author}{\bibfnamefont{X.-G.} \bibnamefont{Wen}},
  \bibinfo{journal}{Advances in Physics} \textbf{\bibinfo{volume}{44}},
  \bibinfo{pages}{405} (\bibinfo{year}{1995}).

\bibitem[{\citenamefont{Haldane and Rezayi}(1985)}]{HR8529}
\bibinfo{author}{\bibfnamefont{F.~D.~M.} \bibnamefont{Haldane}}
  \bibnamefont{and} \bibinfo{author}{\bibfnamefont{E.~H.}
  \bibnamefont{Rezayi}}, \bibinfo{journal}{Phys. Rev. B}
  \textbf{\bibinfo{volume}{31}}, \bibinfo{pages}{2529} (\bibinfo{year}{1985}).

\bibitem[{\citenamefont{Wen and Niu}(1990)}]{WNtop}
\bibinfo{author}{\bibfnamefont{X.-G.} \bibnamefont{Wen}} \bibnamefont{and}
  \bibinfo{author}{\bibfnamefont{Q.}~\bibnamefont{Niu}},
  \bibinfo{journal}{Phys. Rev. B} \textbf{\bibinfo{volume}{41}},
  \bibinfo{pages}{9377} (\bibinfo{year}{1990}).

\bibitem[{\citenamefont{Arovas et~al.}(1984)\citenamefont{Arovas, Schrieffer,
  and Wilczek}}]{ASW8422}
\bibinfo{author}{\bibfnamefont{D.}~\bibnamefont{Arovas}},
  \bibinfo{author}{\bibfnamefont{J.~R.} \bibnamefont{Schrieffer}},
  \bibnamefont{and} \bibinfo{author}{\bibfnamefont{F.}~\bibnamefont{Wilczek}},
  \bibinfo{journal}{Phys. Rev. Lett.} \textbf{\bibinfo{volume}{53}},
  \bibinfo{pages}{722} (\bibinfo{year}{1984}).

\bibitem[{\citenamefont{Halperin}(1982)}]{H8285}
\bibinfo{author}{\bibfnamefont{B.~I.} \bibnamefont{Halperin}},
  \bibinfo{journal}{Phys. Rev. B} \textbf{\bibinfo{volume}{25}},
  \bibinfo{pages}{2185} (\bibinfo{year}{1982}).

\bibitem[{\citenamefont{Wen}(1992)}]{Wedgerev}
\bibinfo{author}{\bibfnamefont{X.-G.} \bibnamefont{Wen}},
  \bibinfo{journal}{Int. J. Mod. Phys. B} \textbf{\bibinfo{volume}{6}},
  \bibinfo{pages}{1711} (\bibinfo{year}{1992}).

\bibitem[{\citenamefont{Wen and Wu}(1993)}]{WWtran}
\bibinfo{author}{\bibfnamefont{X.-G.} \bibnamefont{Wen}} \bibnamefont{and}
  \bibinfo{author}{\bibfnamefont{Y.-S.} \bibnamefont{Wu}},
  \bibinfo{journal}{Phys. Rev. Lett.} \textbf{\bibinfo{volume}{70}},
  \bibinfo{pages}{1501} (\bibinfo{year}{1993}).

\bibitem[{\citenamefont{Chen et~al.}(1993)\citenamefont{Chen, Fisher, and
  Wu}}]{CFW9349}
\bibinfo{author}{\bibfnamefont{W.}~\bibnamefont{Chen}},
  \bibinfo{author}{\bibfnamefont{M.~P.~A.} \bibnamefont{Fisher}},
  \bibnamefont{and} \bibinfo{author}{\bibfnamefont{Y.-S.} \bibnamefont{Wu}},
  \bibinfo{journal}{Phys. Rev. B} \textbf{\bibinfo{volume}{48}},
  \bibinfo{pages}{13749} (\bibinfo{year}{1993}).

\bibitem[{\citenamefont{Senthil et~al.}(1999)\citenamefont{Senthil, Marston,
  and Fisher}}]{SMF9945}
\bibinfo{author}{\bibfnamefont{T.}~\bibnamefont{Senthil}},
  \bibinfo{author}{\bibfnamefont{J.~B.} \bibnamefont{Marston}},
  \bibnamefont{and} \bibinfo{author}{\bibfnamefont{M.~P.~A.}
  \bibnamefont{Fisher}}, \bibinfo{journal}{Phys. Rev. B}
  \textbf{\bibinfo{volume}{60}}, \bibinfo{pages}{4245} (\bibinfo{year}{1999}).

\bibitem[{\citenamefont{Read and Green}(2000)}]{RG0067}
\bibinfo{author}{\bibfnamefont{N.}~\bibnamefont{Read}} \bibnamefont{and}
  \bibinfo{author}{\bibfnamefont{D.}~\bibnamefont{Green}},
  \bibinfo{journal}{Phys. Rev. B} \textbf{\bibinfo{volume}{61}},
  \bibinfo{pages}{10267} (\bibinfo{year}{2000}).

\bibitem[{\citenamefont{Wen}(2000)}]{Wctpt}
\bibinfo{author}{\bibfnamefont{X.-G.} \bibnamefont{Wen}},
  \bibinfo{journal}{Phys. Rev. Lett.} \textbf{\bibinfo{volume}{84}},
  \bibinfo{pages}{3950} (\bibinfo{year}{2000}).

\bibitem[{\citenamefont{Senthil et~al.}(2004)\citenamefont{Senthil, Vishwanath,
  Balents, Sachdev, and Fisher}}]{SVB0490}
\bibinfo{author}{\bibfnamefont{T.}~\bibnamefont{Senthil}},
  \bibinfo{author}{\bibfnamefont{A.}~\bibnamefont{Vishwanath}},
  \bibinfo{author}{\bibfnamefont{L.}~\bibnamefont{Balents}},
  \bibinfo{author}{\bibfnamefont{S.}~\bibnamefont{Sachdev}}, \bibnamefont{and}
  \bibinfo{author}{\bibfnamefont{M.~P.~A.} \bibnamefont{Fisher}},
  \bibinfo{journal}{Science} \textbf{\bibinfo{volume}{303}},
  \bibinfo{pages}{1490} (\bibinfo{year}{2004}).

\bibitem[{\citenamefont{Levin and Wen}(2005)}]{LWstrnet}
\bibinfo{author}{\bibfnamefont{M.}~\bibnamefont{Levin}} \bibnamefont{and}
  \bibinfo{author}{\bibfnamefont{X.-G.} \bibnamefont{Wen}},
  \bibinfo{journal}{Phys. Rev. B} \textbf{\bibinfo{volume}{71}},
  \bibinfo{pages}{045110} (\bibinfo{year}{2005}).

\bibitem[{\citenamefont{Wen}(2004)}]{Wen04}
\bibinfo{author}{\bibfnamefont{X.-G.} \bibnamefont{Wen}},
  \emph{\bibinfo{title}{Quantum Field Theory of Many-Body Systems -- From the
  Origin of Sound to an Origin of Light and Electrons}}
  (\bibinfo{publisher}{Oxford Univ. Press}, \bibinfo{address}{Oxford},
  \bibinfo{year}{2004}).

\bibitem[{\citenamefont{Affleck et~al.}(1988)\citenamefont{Affleck, Zou, Hsu,
  and Anderson}}]{AZH8845}
\bibinfo{author}{\bibfnamefont{I.}~\bibnamefont{Affleck}},
  \bibinfo{author}{\bibfnamefont{Z.}~\bibnamefont{Zou}},
  \bibinfo{author}{\bibfnamefont{T.}~\bibnamefont{Hsu}}, \bibnamefont{and}
  \bibinfo{author}{\bibfnamefont{P.~W.} \bibnamefont{Anderson}},
  \bibinfo{journal}{Phys. Rev. B} \textbf{\bibinfo{volume}{38}},
  \bibinfo{pages}{745} (\bibinfo{year}{1988}).

\bibitem[{\citenamefont{Dagotto et~al.}(1988)\citenamefont{Dagotto, Fradkin,
  and Moreo}}]{DFM8826}
\bibinfo{author}{\bibfnamefont{E.}~\bibnamefont{Dagotto}},
  \bibinfo{author}{\bibfnamefont{E.}~\bibnamefont{Fradkin}}, \bibnamefont{and}
  \bibinfo{author}{\bibfnamefont{A.}~\bibnamefont{Moreo}},
  \bibinfo{journal}{Phys. Rev. B} \textbf{\bibinfo{volume}{38}},
  \bibinfo{pages}{2926} (\bibinfo{year}{1988}).

\bibitem[{\citenamefont{Wen et~al.}(1989)\citenamefont{Wen, Wilczek, and
  Zee}}]{WWZcsp}
\bibinfo{author}{\bibfnamefont{X.-G.} \bibnamefont{Wen}},
  \bibinfo{author}{\bibfnamefont{F.}~\bibnamefont{Wilczek}}, \bibnamefont{and}
  \bibinfo{author}{\bibfnamefont{A.}~\bibnamefont{Zee}},
  \bibinfo{journal}{Phys. Rev. B} \textbf{\bibinfo{volume}{39}},
  \bibinfo{pages}{11413} (\bibinfo{year}{1989}).

\bibitem[{\citenamefont{Ran and Wen}(2005)}]{RWtoap}
\bibinfo{author}{\bibfnamefont{Y.}~\bibnamefont{Ran}} \bibnamefont{and}
  \bibinfo{author}{\bibfnamefont{X.-G.} \bibnamefont{Wen}},
  \bibinfo{journal}{to appear}  (\bibinfo{year}{2005}).

\bibitem[{\citenamefont{Wen}(2002)}]{Wqoslpub}
\bibinfo{author}{\bibfnamefont{X.-G.} \bibnamefont{Wen}},
  \bibinfo{journal}{Phys. Rev. B} \textbf{\bibinfo{volume}{65}},
  \bibinfo{pages}{165113} (\bibinfo{year}{2002}).

\bibitem[{\citenamefont{Wen and Zee}(2002)}]{WZqoind}
\bibinfo{author}{\bibfnamefont{X.-G.} \bibnamefont{Wen}} \bibnamefont{and}
  \bibinfo{author}{\bibfnamefont{A.}~\bibnamefont{Zee}},
  \bibinfo{journal}{Phys. Rev. B} \textbf{\bibinfo{volume}{66}},
  \bibinfo{pages}{235110} (\bibinfo{year}{2002}).

\end{thebibliography}
\end{document}